\documentstyle[11pt,epsfig]{article}
\def\cite#1{#1}
\newcommand{\ct}[1]{[\cite{#1}]}
\def\thebibliography#1{\section*{References}\list
 {[\arabic{enumi}]}{\settowidth\labelwidth{[#1]}\leftmargin\labelwidth
 \advance\leftmargin\labelsep
 \usecounter{enumi}}
 \def\newblock{\hskip .11em plus .33em minus -.07em}
 \sloppy
 \sfcode`\.=1000\relax}

\begin{document}

\def\sa{\omega}
\def\sbb{{\omega^{'}}}
\def\sc{{\omega^{''}}}
\def\sd{{\phi}}
\def\se{{\phi^{'}}}
\def\va{{\varrho}}
\def\vb{{\varrho^{'}}}
\def\vc{{\varrho^{''}}}
\def\vd{{\varrho^{'''}}}
\def\xxx#1#2{\frac{(#1_N-#1_#2)(#1_N-#1_#2^*)(#1_N-1/#1_#2)(#1_N-1/#1_#2^*)}
               {(#1-#1_#2)(#1-#1_#2^*)(#1-1/#1_#2)(#1-1/#1_#2^*)}}
\def\eee#1#2{\frac{(#1_N-#1_#2)(#1_N-#1_#2^*)(#1_N+#1_#2)(#1_N+#1_#2^*)}
               {(#1-#1_#2)(#1-#1_#2^*)(#1+#1_#2)(#1+#1_#2^*)}}
\def\fff#1#2{(f_{#2{KK}}/f_{#2}^e)}
\def\ff#1#2{(f_{#2{KK}}/f_{#2}^s)}

\begin{center}
 {\Large\bf STRANGE  VECTOR  FORM FACTOR OF KAONS}
\end{center}
\begin{center}
{S. Dubni\v cka}        \\
{\em Inst. of Physics, Slovak Academy of Sciences, D\'ubravsk\'a cesta 9,
842 28 Bratislava, Slovak Republic}\\
{A. Z. Dubni\v ckov\'a}   \\
{\em Dept. of Theoretical Phys., Comenius University, Mlynsk\'a dolina, 842
48 Bratislava, Slovak Republic}\\
\end{center}
\begin{abstract}
Starting from the $\omega-\phi$ mixing, further assuming the coupling
of the quark-current of some flavour  to be of universal strenth
exclusively to the component of vector-meson wave function with
the same flavour and finally taking numerical values of the
coupling constant ratios ($f_{\omega K\bar K}/f_{\omega}^e$),
($f_{\phi K\bar K}/f_{\phi}^e$) from the isoscalar part of a realistic 
six-resonance unitary and analytic model of the kaon 
electromagnetic structure, the strange-quark vector current form 
factor behaviour of K-mesons in space-like and time-like regions 
is predicted.
\end{abstract}

Indirect experimental evidence for strangeness in nucleon in the 
determination of the $\pi N$ $\sigma$-term, in polarized deep-inelastic 
lepton-nucleon scattering, in violations of the OZI rule, in elastic
$\nu p$ scattering and in neutrino charm production motivated in recent
years, among others, also a realization of well-defined experimental \ct{1-5}
and theoretical \ct{6-30} program of an investigation of the matrix 
element $\langle N|\bar s\gamma_{\mu} s|N\rangle$, which is directly
related to the strange nucleon electric and magnetic form factors (ff's),
$G_E^{s}$ and $G_M^{s}$. Although the present experimental values \ct{1-5}
of the strange nucleon ff's are almost consistent with zero, the rather
sizeable error bars document the difficulty of such type experiments
and no definitive conclusion can as yet be made regarding the experimental 
scale of $\langle N|\bar s\gamma_{\mu} s|N\rangle$. On the other hand the 
theoretical understanding of $\langle N|\bar s\gamma_{\mu} s|N\rangle$
is even much less clear. There have been many theoretical attempts
to predict the nucleon strange form factors, differing by the methods used
to study the non-perturbative physics behind them: VMD models with 
$\omega-\phi$ mixing \ct{6,10,15,19}, Skyrme models \ct{7,9},
Nambu-Jona-Lasinio soliton model \ct{14}, SU(3) chiral bag model \ct{20},
kaon loop model \ct{8,11,13,17}, meson exchange model \ct{16}, lattice
QCD \ct{21,26,27,28}, chiral quark model \ct{25}, SU(3) chiral quark-soliton
model \ct{30}, HBChPT \ct{22,24}, unitary and analytic model with 
$\omega-\phi$ mixing \ct{29} and dispersion relation approach \ct{23}.
So, the range of predictions for the strange nucleon ff's is broad and
obtained results are mutulally contradicting. The
latter illustrates the sensitivity of see quark observables to model 
assumptions  and, on the other hand, the limited usefulness of models 
in making realistic predictions.
One could hope for more insight from ChPT, which relies on the chiral symmetry 
of QCD. However, the application of ChPT is restricted in energy and mereover,
depends \ct{24} on unknown counterterms, which have to be determined
from data, including the data on the strange nucleon ff's to be
charged by sizable error bars, for which one expects predictions just 
from ChPT.
In this paper we turn our attention to the dispersion relation approach
\ct{23}, which, similarly to ChPT, relies on general principles of quantum field
theory, including also QCD, and seems to bring some insight into 
$\langle N|\bar s\gamma_{\mu} s|N\rangle$ as well, relating existing 
experimental data to the observables of interest. In the dispersion
relation approach it is analyticity and unitarity, rather than
chiral symmetry, which allow one to make such a relation. Practically
it means first, on the base of the assumed analytic properties and the
asymptotic behaviours of ff's, to write down substracted (or unsubracted)
dispersion integral relations and then to use the unitarity  condition
for a derivation of relations of the imaginary parts of ff's under the 
integrals to amplitudes of physical processes. Though, there is an
infinit number of contributing terms into imaginary parts of ff's, the
lightest intermediate states in the unitarity condition generate dominant
contributions to the leading moments of the strange current. The lowest
one is the 3$\pi$ contribution, which can resonate into a state having 
the same quantum numbers as
the $\phi$-meson (nearly pure $s\bar s$ state).
Thus, the 3$\pi$ state can contribute  appreciably to the strange nucleon
ff's via its coupling to the $\phi$-meson. The next important is the
$K\bar K$ intermediate state, which is the lightest state containing 
valence strange quarks and represents so-called ``kaon cloud dominance''
in the strange nucleon ff's.

In order to estimate $K\bar K$ contributions to the corresponding spectral
functions, following the analysis in \ct{31}, one can express e.g. the 
absorptive part of the strange nucleon electric ff $G_E^{s}(t)$ (and similarly
for $G_M^{s}(t)$) as a product
of the appropriate  J=1 $K \bar K \to N \bar N$ partial wave amplitude 
$b_1^{\lambda, \lambda{'}}$ and the strange vector form factor of kaons 
$F_K^{s}(t)$ as follows \ct{32}
\begin{equation}
Im G_E^{s}= Re \left \{\left( \frac{q}{4m_N}\right )b_1^{1/2,1/2}(t)\cdot F_K^{s}(t)^*\right \},
\label{r1}
\end{equation}
where $q=\sqrt{t/4-m_K^2}$ and $\lambda, \lambda{'}$ in $b_1^{\lambda, \lambda{'}}$
are denoting corresponding helicities.

In the spirit of a discussion above, that futher theoretical improvements
are needed to constrain the uncertainties in the experimental measurements
of the strange nucleon ff's, the problem is now to determine $b_1^{\lambda,
\lambda{'}}$
and $F_K^{s}(t)$ as reliably as possible.

Whereas the $K \bar K \to N\bar N$ J=1 partial wave  $b_1^{1/2,1/2}$
can be determined \ct{31,32} by an analytic continuation of the data on the
$K^+ N\to K^+ N$ amplitude into the unphysical region, there exists no experimental
infomation on $F_K^{s}(t)$ and one has to find a position in some specific
models of $F_K^{s}(t)$, like a simple $\phi$-dominance form \ct{31},
Gounaris-Sakurai parametrization \ct{33} of the electromagnetic (EM) ff
  $F_{\pi}^{EM}(t)$ with the replacement 
of the $\rho$ mass and width with those of $\phi$, or the sum of Breit-Wigner terms
\ct{23} of the form \ct{34}
\begin{equation}
F_K^{s}=\sum_{v=\omega,\phi} C_v^{s}\frac{m^2_v}{m^2_v-t-im_v\Gamma_v f_v(t)}
\label{r2}
\end
{equation}
where $f_v(t)=t/m^2_v$ and $C_v^{s}$ are determined by the relations \ct{6}
\begin{eqnarray}
C_{\omega}^{s}/C_{\omega}^{e} &\sim& -0.2 \nonumber\\
C_{\phi}^{s}/C_{\phi}^{e} &\sim& -3\label{r3}
\end{eqnarray}
from the values of $C_{\omega}^e$, $C_{\phi}^e$ determined in a fit of data on
the kaon EM ff's measured by $e^+e^-\to K\bar K$ processes.

In this paper we follow the procedure in \ct{23}, but
 for a parametrization of the kaon EM ff's and kaon strange ff 
we apply more sophisticated unitary and analytic models, which unify all
known ff properties always into one analytic function in a very natural way
and as a result, one can expect to obtaine  realistic behaviour of $F_K^s(t)$.

The strange-quark vector current ff of K-mesons 
$ F_{K}^{s}(t)$ is defined by the matrix element of the strange-quark current 
$J_{\mu}^{s}= \bar s\gamma_{\mu} s$
\begin{equation}
\langle {K}({k'})|\bar {\rm s}\gamma_{\mu} {\rm s}| {K}({k})\rangle =
(k+k^{'})_{\mu} {F}_{K}^{s}(t)
\label{rov1}
\end{equation}
in analogy to a definition of the EM ff's of K -mesons, $F_{{K^+}}({t})$ 
and $F_{{K^0}}({t})$
\begin{equation}
\langle {K}({k'})|J_{\mu}^{EM}| {K}({k})\rangle = ({k} + {k'})_{\mu}
{F}_{K}(t)
\label{rov2}
\end{equation}
where
\begin{equation}
 J_{\mu}^{{\rm EM}}  = \frac{2}{3}\bar {\rm u}\gamma_{\mu} {\rm u}- \frac{1}{3}\bar {\rm d}\gamma_{\mu} {\rm d} -\frac{1}{3}\bar {\rm s}\gamma_{\mu} {\rm s}
\label{rov3}
\end{equation}
is the EM current operator written by means of the ${\rm u}$, ${\rm d}$ and
${s}$ quark fields, $k$ and $k'$ are four momenta of K-mesons and
${t}=(k' - k)^2=  q^2=  -Q^2$ is the four momentum
transfer squared.

The ff's $F_{{K^+}}(t)$ and $F_{{K^0}}(t)$
can be decomposed into isoscalar and isovector parts
\begin{eqnarray}
 F_{K^+}&=& F_{K}^{I=0}(t) + F_K^{I=1}(t) \nonumber \\
 F_{K^0}&=& F_{K}^{I=0}(t) -  F_K^{I=1}(t) \label{rov4}
\end{eqnarray}
to be defined by the following matrix elements
\begin{eqnarray}
\langle {K}({k'})|J_{\mu}^{I=0} | {K}({k})\rangle &=& ({k} + {k'})_{\mu} {F}_{K}^{I=0}(t) \nonumber \\
\langle {K}({k'})|J_{\mu}^{I=1} | {K}({k})\rangle &=& ({k} + {k'})_{\mu} {F}_{K}^{I=1}(t), \label{rov5}
\end{eqnarray}
where  $J_{\mu}^{I=0}  = \frac{1}{6}(\bar {\rm u}\gamma_{\mu} u + \bar {\rm d}\gamma_{\mu} {\rm d}) -\frac{1}{3}\bar {\rm s}\gamma_{\mu} {\rm s}$
and $J_{\mu}^{I=1}  = \frac{1}{2}(\bar {\rm u}\gamma_{\mu} u-\bar {\rm d}\gamma_{\mu} {\rm d})$
are isoscalar and isovector parts of the EM current $ J_{\mu}^{{\rm EM}}$
given by (\ref{rov3}), respectively.

Since the strange-quark vector current $J_{\mu}^{s}  =  \bar {\rm s}\gamma_{\mu} {\rm s}$
carries the quantum numbers of the isoscalar part of the EM current
$ J_{\mu}^{I=0}$, then the strange ff  $F_{K}^{s}(t)$ can
contribute only to a behaviour of the isoscalar part
$F_{K}^{I=0}(t)$ of the kaon EM ff's. So, it is
natural to expect that in principle one  could draw out the behaviour of
$F_{K}^{s}(t)$ just from the isoscalar part
$F_{K}^{I=0}(t)$ of the kaon EM ff's. Really, if there is a
suitable model of
\begin{equation}
F_K^{I=0}(t)= {\huge f}[t; (f_{\omega K K}/f_{\omega}^e), (f_{\phi K K}/f_{\phi}^e)]
\label{rov6}
\end{equation}
and
\begin{equation}
F_{K}^{s}(t)= {\tilde{\huge f}}[t; (f_{\omega K K}/f_{\omega}^s), (f_{\phi K K}/f_{\phi}^s)]
\label{rov7}
\end{equation}
is the model of the same analytic structure, but with different norm and
different asymptotic behaviour (therefore denoted by $\tilde {\huge f}$), then starting
from the $\omega$-$\phi$ mixing and assuming that the quark-current of some
flavour couples with universal strength exclusively to the component of the
vector-meson wave function with the same flavour, one can prove \ct{6}
the following relations
\begin{eqnarray}
(f_{\omega K K}/f_{\omega}^s)&=& -{\sqrt{6}}\frac{\sin{\epsilon}}{\sin{(\epsilon+\theta_0)}}(f_{\omega K K}/f_{\omega}^e);
\label{rov8}\\
(f_{\phi K K}/f_{\phi}^s)&=&- {\sqrt{6}}\frac{\cos{(\epsilon)}}{\cos{(\epsilon+\theta_0)}}(f_{\phi K K}/f_{\phi}^e)\nonumber
\end{eqnarray}

between parameters of the models (\ref{rov7}) and (\ref{rov6}), where
$\epsilon = 3.7^{\circ}$ is a deviation from the ideally mixing angle
$\theta_0=35.3^{\circ}$, and as  a result the behaviour of $F_{{K}}^s(t)$ can be
predicted.

Practically the EM structure of K-mesons we describe by the unitary and analytic
model \ct{35,36}
\begin{eqnarray}
F^{I=0}_K [V(t)]&=&\left (\frac{1-V^2}{1-V^2_N}\right)^2 \left [\frac{1}{2}\eee V\se \right. + \label{rov9} \\
\nonumber &+&\left \{\xxx V\sa \right. - \\
\nonumber &-&\left. \eee V\se \right \}\fff 1\sa+ \\
\nonumber &+&\left \{\xxx V\sd \right. -  \\
\nonumber &-&\left.\left. \eee V\se \right \}\fff 1\sd \right ]
\end{eqnarray}
 \begin{eqnarray}
 \nonumber F^{I=1}_K [W(t)]&=&\left(\frac{1-W^2}{1-W_N^2}\right )^2\left [\frac{1}{2}\eee W\vd \right.+\\
  &+&\left \{\xxx W\va \right.- \label{rov10} \\
\nonumber &-&\left. \eee W\vd \right \}\fff 1\va +   \\
 \nonumber &+&\left\{ \xxx W\vb \right. - \\
\nonumber &-&\left.\left. \eee W\vd \right \}\fff 1\vb \right ],
\end{eqnarray}
where
$$
V(t)=i\frac{\sqrt {q_{in}^{I=0}+ q}-\sqrt{q_{in}^{I=0}-q}}
{\sqrt{q_{in}^{I=0}+q}+\sqrt{q_{in}^{I=0}-q}}
$$
$$
q=[(t-t_0^{I=0})/t_0^{I=0}]^{1/2}; \;\;\; q_{in}=[(t_{in}-t_0^{I=0})/t_0^{I=0}]^{1/2}; \;\;\; V_N=V(t)_{|_{t=0}};
$$
$t_0^{I=0}= 9m_{\pi}^2$,  $t_{in}^{I=0}$ is an effective square-root branch
point simulating contributions of all higher thresholds given by the unitarity
conditions, $V_i$ $(i=\omega, \phi, \phi')$ are the positions of vector-meson poles
in $V(t)$-plane and similarly for $W(t)$, $W_N$ and $W_j$ $(j=\varrho, \varrho', \varrho''')$.

The ff's $F_K^{I=0}[V(t)]$ and $F_K^{I=1}[W(t)]$ reflect all known theoretical
properties of the kaon EM ff's.

Similarly, we construct the unitary and analytic model of the strange-quark
vector current ff of K-mesons
\begin{eqnarray}
 F^{s}_K[V(t)]&=&\left(\frac{1-V^2}{1-V_N^2}\right)^6 \left [-\eee V\se\right. +\label{rov11} \\
\nonumber &+&\left \{\xxx V\sa \right. -\\
\nonumber &-&\left. \eee V\se \right \}\ff 1\sa + \\
\nonumber &+&\left\{\xxx V\sd \right. -\\
\nonumber &-&\left.\left. \eee V\se \right \}\ff 1\sd \right ]
\end{eqnarray}

\begin{figure}[t]
\centerline{\epsfig{file=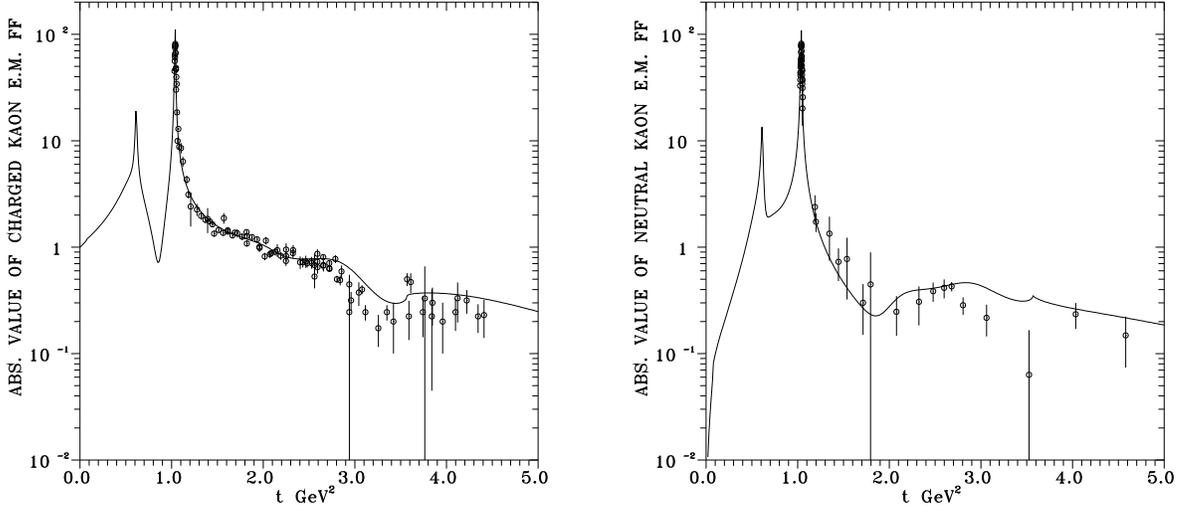,height=7.5cm}}
\caption{
Description of data on the charge and neutral kaon EM ff's by unitary and analytic models.
}
\label{fig.1}
\end{figure}

\begin{figure}[htb]
\centerline{\epsfig{file=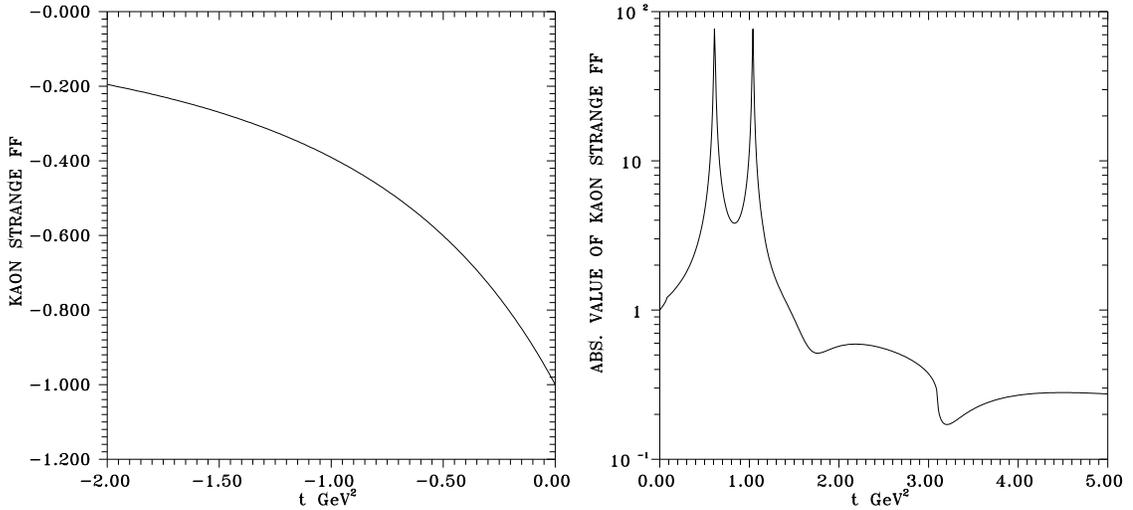,height=7.5cm}}
\caption{Prediction of the strange-quark vector current ff of K-mesons in the space-like
and  the time-like regions.}
\label{fig.2}
\end{figure}


with the same analytic structure as $F_K^{I=0}[V(t)]$, but to be normalized 
to -1, in order to take the correct strangeness charge into account, and
with the asymptotic behaviour $F_K^s(t)_{|t|\to\infty}\sim t^{-3}$, 
as in addition to the valence ($q\bar q$) of K-mesons, now also
the $s\bar s$ - pairs contribute and as a result there is totally 4 quarks 
inside the kaons. The $f_{\omega}^s$ and $f_{\phi}^s$  are
strangeness-current-vector-meson ($\omega$ and $\phi$, respectively)
coupling constants.
\begin{figure}[htb]
\centerline{\epsfig{file=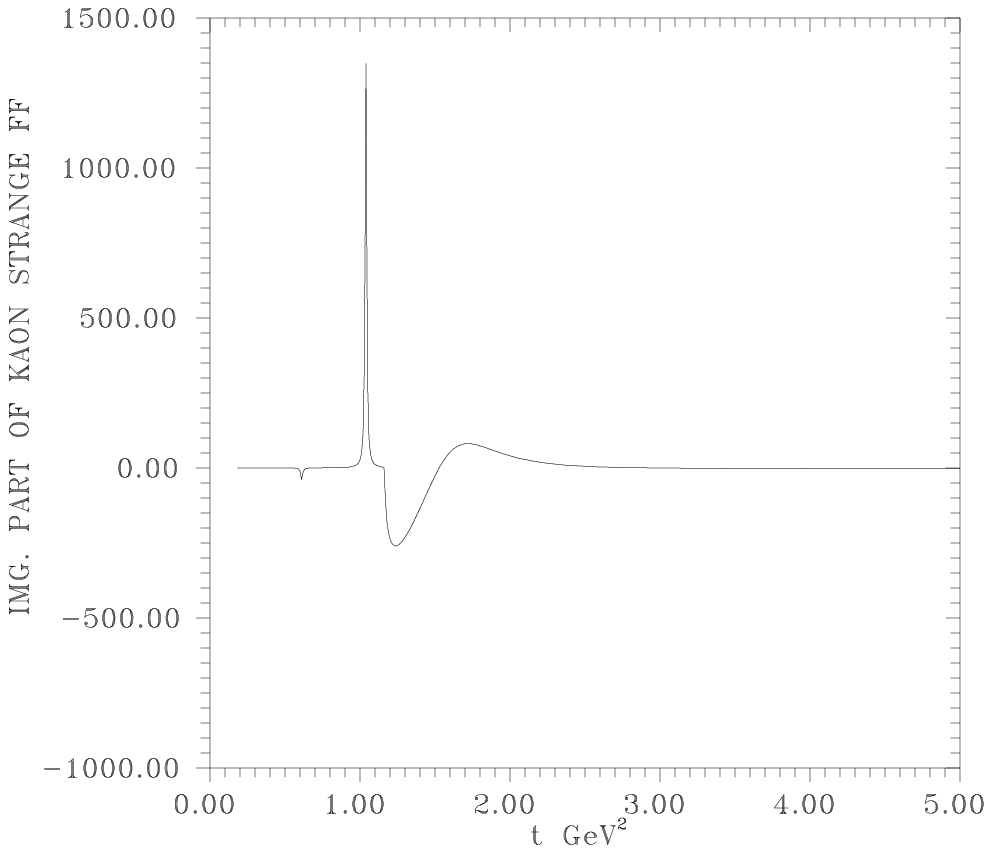,height=7.5cm}}
\caption{The predicted spectral function behaviour of $F_K^s(t)$.}
\label{fig.3}
\end{figure}

With the aim of a determination of free parameters in isoscalar
and isovector parts of EM ff's of K- mesons, $F_K^{I=0}[V(t)]$ and
 $F_K^{I=1}[W(t)]$, the latter are compared through the relations 
(\ref{rov4}) with all existing data on $F_{K^+}$ and $F_{K^0}$,                                                                                                                                      
simultaneously. 
The results with $\chi^2/ndf$=1.6 are graphically presented in Fig.1.
Important   parameters of $F_K^{I=0}[V(t)]$ for a prediction of 
$F_K^{s}[V(t)]$ behaviour are found to be 
\begin{equation}
t_{in}^{I=0}\equiv t_{in}^s=1.0480\pm 0.2800 {\rm GeV}^2
\end{equation}
$$
(f_{\omega K K}/f_{\omega}^e)=0.2076\pm 0.0542; \;\;\;\; (f_{\phi K K}/f_{\phi}^e)=0.3466\pm 0.0491
$$
from where, by means of the relations (\ref{rov8}), one finally evaluates
unknown parameters of $F_K^{s}[V(t)]$ 
in (\ref{rov7}) to be
\begin{equation}
(f_{\omega K K}/f_{\omega}^s)=-0.0522; \;\;\;\;\; (f_{\phi K K}/f_{\phi}^s)=-1.0902.
\end{equation}
                                                                                                                                                                                                                              
Then by means of the unitary and analytic model (\ref{rov11})
the strange-quark vector current
form factor behaviour of K-mesons in space-like and time-like regions, as
graphically presented in Fig.2, is predicted which is formed also 
by $\phi$-meson, similarly to \ct{23},  but besides the latter also
$\omega$-meson contributes and the height and the shape of $F_K^s(t)$ are different.

A prediction of a corresponding spectral function behavior is shown in Fig.3.
Here we would like to note, that the obtained results depend on the precision of
existing data on the kaon EM ff's.
One hopes that the substantial improvement, at least at the $\phi$-region, will be
obtained by VEPP-2M and DA$\Phi$NE soon. Consequently even more accurate predictions for
$F_K^s[V(t)]$ behaviour could be achieved.

One of the authors (S.D.) wish to thank  Mauro Anselmino for stimulating
suggestions and fruitful discussions during his stay at the Torino University, where
the part of this work was done.

The work was in part supported by the Slovak Grant Agency for Sciences,
Gr. No. 2/5085/2000 (S.D.) and Gr. No. 1/7068/2000 (A.Z.D.).


\begin{thebibliography}{99}
\bibitem{1}
B. Mueller et al., Phys. Rev. Lett. {\bf 78} (1997) 3824.
\bibitem{2}
K. Aniol et al.,  Phys. Rev. Lett. {\bf 82} (1999) 1096.
\bibitem{3}
D.T. Spayde et al.,  Phys. Rev. Lett. {\bf 84} (2000) 1106.
\bibitem{4}
R. Hasty et al., Science {\bf 290} (2000) 2117.
\bibitem{5}
K.A. Aniol et al., Phys. Lett. {\bf B509} (2001) 211.
 \bibitem{6}
R.L.~Jaffe, Phys. Lett. {\bf B229} (1989) 275.
\bibitem{7}
N.V. Park, J. Schechter, H. Weigel, Phys. Rev. {\bf D43} (1991) 869.
\bibitem{8}
W. Koepf, E.M. Henley, J.S. Pollock, Phys. Lett. {\bf B288} (1992) 11.
\bibitem{9}
N.W. Park, H. Weigel, Nucl. Phys {\bf A541} (1992) 453.
\bibitem{10}
H. Forkel, M. Nielsen, X. Jin, T.D. Cohen, Phys. Rev. {\bf C50} (1994) 3108.
\bibitem{11}
M.J. Musolf, M. Burkhard, Z. Phys. {\bf C61} (1994) 433.
\bibitem{12}
M.J. Musolf et al., Phys. Rep. {\bf 239} (1994) 1.
\bibitem{13}
H. Ito, Phys. Rev {\bf C52} (1995) R1750.
\bibitem{14}
H. Weigel et al., Phys. Lett. {\bf B353} (1995) 20.
\bibitem{15}
H.-W. Hammer, U.-G. Meissner, D. Drechsel, Phys. Lett. {\bf B367} (1996) 323.
\bibitem{16}
 U.-G. Meissner, V. Mull, J. Speth, J.W. Van Orden, Phys. Lett. {\bf B408} (1997) 381.
\bibitem{17}
P. Geiger, N. Isgur, Phys. Rev. {\bf D55} (1997) 299.
\bibitem{18}
B.-Q. Ma, Phys. Lett. {\bf B408} (1997) 387.
\bibitem{19}
H. Forkel, Phys. Rev. {\bf C56} (1997) 510.
\bibitem{20}
S.-T. Hong, B.-Y. Park, D.-P. Min, Phys. Lett. {\bf B414} (1997) 229.
\bibitem{21}
S.J. Dong, K.-F. Lin, A.G. Wiliams, Phys. Rev. {\bf D58} (1998) 074504.
\bibitem{22}
T.R. Hemmert, U.-G. Meissner, S. Steininger, Phys. Lett. {\bf B437} (1998) 184. 
\bibitem{23}
H.-W. Hammer, M.J. Ramsey-Musolf, Phys. Rev.{\bf C60} (1999) 045205.
\bibitem{24}
T.R. Hemmert, B. Kubis, U.-G. Meissner, Phys. Rev. {\bf C60} (1999)
045501. 
\bibitem{25}
L. Hannelius, D.O. Riska, L.Ya. Glozman, Nucl. Phys. {\bf A665} (2000) 353.
\bibitem{26}
K.-F. Lin, hep-ph/0011225.
\bibitem{27}
D.B. Leinweber, A.W. Thomas, Phys. Rev. {\bf D62} (2000) 074505.
\bibitem{28}
N. Mathur, S.-J. Dong, Nucl. Phys. Proc. Supl. {\bf 94} (2001) 311.
\bibitem{29}
S.~Dubni\v cka, A.Z.~Dubni\v ckov\'a, P.~Weisenpacher, hep-ph/0102171.
\bibitem{30}
A. Silva, H.-Ch. Kim, K. Gocke, hep-ph/0107185.
\bibitem{31}
M.J. Musolf, H.-W. Hammer, D. Drechsel, Phys. Rev. {\bf D55} (1997) 2741.
\bibitem{32}
 M.J. Ramsey-Musolf, H.-W. Hammer, Phys. Rev.  Lett. {\bf 80} (1998) 2539. 
\bibitem{33}
G.J. Gounaris, J.J. Sakurai, Phys. Rev. Lett. {\bf 21} (1968) 244.
\bibitem{34}
F. Felicetti, Y. Srivastava, Phys. Lett. {\bf B107} (1981) 227.
\bibitem{35}
S.~Dubni\v cka, A.Z.~Dubni\v ckov\'a, P. Str\'{\i}\v zenec, hep-ph/0108053.
\bibitem{36}
S.~Dubni\v cka, Report JINR, E2-88-840, Dubna, (1988).
\end{thebibliography}
\end{document}